\documentclass[journal, twoside]{IEEEtran}
\usepackage{amsmath}
\usepackage{amsfonts}
\usepackage{graphicx}
\usepackage{epstopdf}
\usepackage{tikz}
\usepackage{booktabs}

\usepackage[
    colorlinks=true,
    linkcolor=blue,
    urlcolor=blue,
    citecolor=blue,
    bookmarksdepth=3,
    pdfstartview =FitH,
    bookmarksopenlevel	= \maxdimen,
    pdfauthor=Ernesto \ Accolti,
    pdftitle={Preprint: Effect of laboratory conditions on the perception of virtual stages for music},
]{hyperref}

\begin{document}

\title{Effect of laboratory conditions on the perception of virtual stages for music}

 \author{Ernesto Accolti
 \thanks{Ernesto Accolti is with the Instituto de Autom\'atica (Institute of Automation), depending on both the National University of San Juan (UNSJ) and the National Scientific and Technical Research Council (CONICET). Email: eaccolti@inaut.unsj.edu.ar. 
 }}

	

    \maketitle
	 
\begin{abstract}
This manuscript presents initial findings critical for supporting augmented acoustics experiments in custom-made hearing booths, addressing a key challenge in ensuring perceptual validity and experimental rigor in these highly sensitive setups.
This validation ensures our proposed methodology is sound, guarantees the reliability of future results, and lays the foundational groundwork for subsequent perceptual studies and the development of robust guidelines for laboratory design in virtual acoustics research.
A preliminary study on the effect of the acoustical conditions of three different rooms on the perception of virtual stages for music is presented: an anechoic room, a custom-made hearing booth with insufficient sound absorption, and another custom-made hearing booth with achievable sound absorption.
The goal of this study is to assess the impact of these different conditions on the perception of virtual stages for music.
The results show that the anechoic room and the hearing booth with achievable sound absorption have a difference between the total sound and the virtual sound below the just-noticeable difference, which means that the virtual sound is not perceived louder than it should.
In contrast, the hearing booth with insufficient sound absorption has a difference above the just-noticeable difference, which means that the virtual sound is perceived louder than it should.
This study provides a preliminary validation of the proposed methodology for assessing the acoustical conditions of custom-made hearing booths in stage acoustics experiments.
Future work will include a more comprehensive analysis of the results, including the effect of different sound sources.

Supplementary audio files illustrating key simulation results are available at \url{https://zenodo.org/records/15579861}.

\end{abstract}

\section{Introduction}
\label{sec:intro}

This preliminary draft presents initial findings critical for supporting a grant proposal focused on augmented acoustics experiments in custom-made hearing booths. 
Given the high sensitivity of these experiments to the soundfield of the real room, thoroughly assessing the booths' acoustical conditions is essential.
This validation ensures our proposed methodology is sound and guarantees the reliability of future results.
A more comprehensive, peer-reviewed version of this work will follow.

Real time soundfield simulations are frequently used in stage acoustics research. 
Early experiments on stage acoustics were carried out by separating musicians of ensembles in two different anechoic rooms available in the same building, which is a condition not easily meet at laboratory buildings \cite{gade1989}.
Currently, the most common approach is to use a single anechoic room and set up the experiments with all the musicians of the ensemble playing simultaneously in the same room \cite{braak2019}.

While real-time soundfield simulations are increasingly used in stage acoustics, the interaction between the virtual acoustic environment and the unavoidable physical acoustic properties of the laboratory space remains a critical, yet often underestimated, factor affecting experimental validity. 
Early experiments in separate anechoic rooms \cite{gade1989} highlight the ideal, but not always practical, conditions. Modern approaches often rely on single anechoic rooms \cite{braak2019}, yet some institutions, seeking to conduct this research, face the challenge of designing and validating more accessible, custom-made hearing booths. 
This preliminary study directly addresses this gap by systematically evaluating the perceptual impact of different laboratory acoustic conditions on virtual stage perception.

This paper presents a preliminary study on the effect of the acoustical conditions of three different rooms on the perception of virtual stages for music: 
an anechoic room which may be the best scenario, a custom-made hearing booth with not enough sound absorption which may be the worst scenario, and another custom-made hearing booth with achievable sound absorption which may be a compromise scenario.  
The goal of this study is to assess the impact of these different conditions on the perception of virtual stages for music. 

The paper is organized as follows:
Section \ref{sec:methods} describes the materials and methods used in the study, including the description of the laboratory rooms, the virtual stages, the soundfield simulation, and the definition of the binaural virtual sound, the binaural residual sound, and the total sound.
Section \ref{sec:results} presents the preliminary results of the study, including the sound pressure levels and the difference between the total sound and the virtual sound.
Section \ref{sec:discussion} discusses the results, including the signal-to-noise ratio and the just-noticeable difference.
Finally, Section \ref{sec:conclusion} concludes the paper and outlines future work.

\section{Materials and Methods}
\label{sec:methods}

\subsection{Description of the laboratory rooms}
The main characteristics 
of the three laboratory rooms studied in this paper are summarized in Table \ref{tab:rooms}.
For the sake of simplicity, a constant mean absorption coefficient $\overline{\alpha}$ is used to characterize the acoustical conditions of the three laboratory rooms.  It is acknowledged that actual absorption is frequency- and material-dependent, factors that will be explored in more detailed future work.
The conditions can be meet at least for frequencies above 120 Hz for the Hearing Booth 2 by the use of 120 mm rock wool panels with an air cavity of 100 mm. 
Eventually, in case the conditions do not meet in practice, membrane and Helmholtz resonators can be used to fine tune the absorption coefficient of the booths \cite{pausch2022}.   
\begin{table}[h]
\centering
\caption{Main conditions of the three laboratory rooms}
\label{tab:rooms}
\begin{tabular}{c c c c c}
\toprule
\textbf{Room} & \textbf{Width} & \textbf{Length} & \textbf{Height} & $\overline{\alpha}$ \\
\midrule
Anechoic room & 3.5 m & 4.5 m & 2.5 m & 0.99 \\
Hearing booth 1 & 2.0 m & 2.0 m & 2.0 m & 0.50 \\
Hearing booth 2 & 2.1 m & 3.0 m & 2.5 m & 0.97 \\	
\bottomrule
\end{tabular}
\end{table}

\subsection{Description of the virtual stages}
\begin{figure}[h]
\centering
\includegraphics[width=0.40\textwidth]{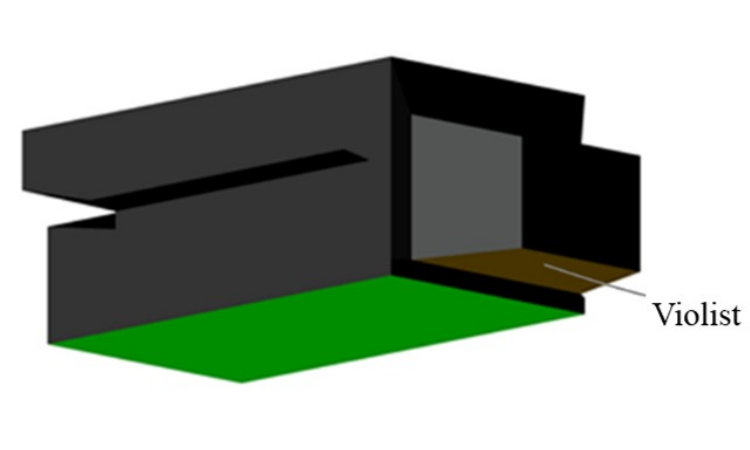}
\caption{Virtual stage model.}
\label{fig:virtual_stage}
\end{figure}

A virtual stage situation is simulated for each of two different stages: a small one with dimensions 12 m width and 6 m height, and a large one with dimensions 24 m width and 12 m height.
Both stages' depth is 10 m. 
The audience's volume is 41.5 m length, 23 m width and 19 m height for both stages. 
The Fig.~\ref{fig:virtual_stage} shows the virtual room model.
The absorption coefficients and scattering coefficients in the central frequency bands are 0.80 and 0.70, respectively, for the audience area, and 0.20 and 0.10 for the rest of the surfaces.

\subsection{Soundfield simulation}
Raven software is used to simulate both the virtual stages and the three laboratory rooms.
The software is based on image source and ray tracing methods \cite{raven}. 
Skipping the direct sound, is a feature of Raven that allows to simulate the sound reflections in a room without the direct sound reaching the listener. 

A violist is placed in the centre of the virtual stages, modelled with the source directivity and head related transfer function from public databases \cite{ackermann2024,brinkmann2017}.
The anechoic sound of the viola is a recording of the first 6 seconds of the third movement of the Summer concerto of the four seasons by Vivaldi (RV315) extracted from the databse of the Sorbonne University \cite{thery2019}.

\subsection{Binaural virtual sound}
The binaural sound the violist is expected to hear $y_\mathrm{v}$ is simulated by the convolution of the anechoic sound $x$ with the left/right ($l/r$) channels of the binaural room impulse response $h_{\mathrm{v}}$ calculated with Raven 
\begin{equation}
\label{eq:yv}
y_{\mathrm{v},l/r} = h_{\mathrm{v},l/r} \ast x.
\end{equation}
In a stage acoustics experiment, $y_{\mathrm{v}}$ has two parts: the direct sound $y_{\mathrm{v,d}}$ which directly reaches the violist, and the sound that is reflected by the walls of the room $y_{\mathrm{v,sk-d}}$ which is modelled by skipping the direct sound in $h_\mathrm{v}$ \cite{accolti2023}.  
\begin{equation}
\label{eq:yv_sk}
y_{\mathrm{v}} = y_{\mathrm{v,d}} + y_{\mathrm{v,sk-d}}
\end{equation}

\subsection{Binaural residual sound} 
Since the musicians in a stage acoustics experiment wear open headphones to hear the direct sound, the sound reflections of the real room also reach the violist. This residual sound $y_\mathrm{u}$ is modelled by the convolution of the anechoic sound with $h_{\mathrm{u}}$ calculated for the laboratory room where the real violist is located but skipping the direct sound which is required to be heard by the violist as defined in \eqref{eq:yv_sk}.  
\begin{equation}
\label{eq:yu}
y_{\mathrm{u},l/r} = h_{\mathrm{l},l/r} \ast x.
\end{equation}
Besides, the total sound reaching the violist ears $y_\mathrm{t}$ is calculated as the sum of the virtual sound and the residual sound
\begin{equation}
\label{eq:yt}
y_{\mathrm{t}} = y_{\mathrm{v}} + y_{\mathrm{u}}.
\end{equation}

\subsection{Sound pressure levels}
The sound pressure levels of the virtual sound $L_\mathrm{v}$, the residual sound $L_\mathrm{u}$, and the total sound $L_\mathrm{t}$ are calculated for each of the three laboratory rooms and for each of the two virtual stages.
A 2-millisecond (ms) moving average was adopted, inspired by psychoacoustic principles, particularly concerning temporal integration within the auditory system. 
This duration, supported by state-of-the-art loudness models \cite{moore2014,chalupper2002}, ensures our estimations accurately capture the rapid amplitude dynamics critical for perceived sound quality in stage acoustics experiments.

In order to compare the ideal situation to the real situation in the studied laboratory rooms, two descriptors are estimated: signal-to-noise ratio (SNR) and difference between total level and ideal level $\Delta$.  
The SNR is defined as
\begin{equation}
\label{eq:snr}
\mathrm{SNR} = L_\mathrm{v} - L_\mathrm{u}.
\end{equation}
Besides, the difference between the total sound and the virtual sound is calculated as
\begin{equation}
\label{eq:deltaL}
\Delta L = L_\mathrm{t} - L_\mathrm{v}.
\end{equation}

\section{Preliminary Results}
\label{sec:results}

\begin{figure*}	
\centering
\begin{tabular}{l l}
a) Small Virtual Hall -- Anechoic room &
b) Large Virtual Hall -- Anechoic room \\
\includegraphics[width=0.45\textwidth]{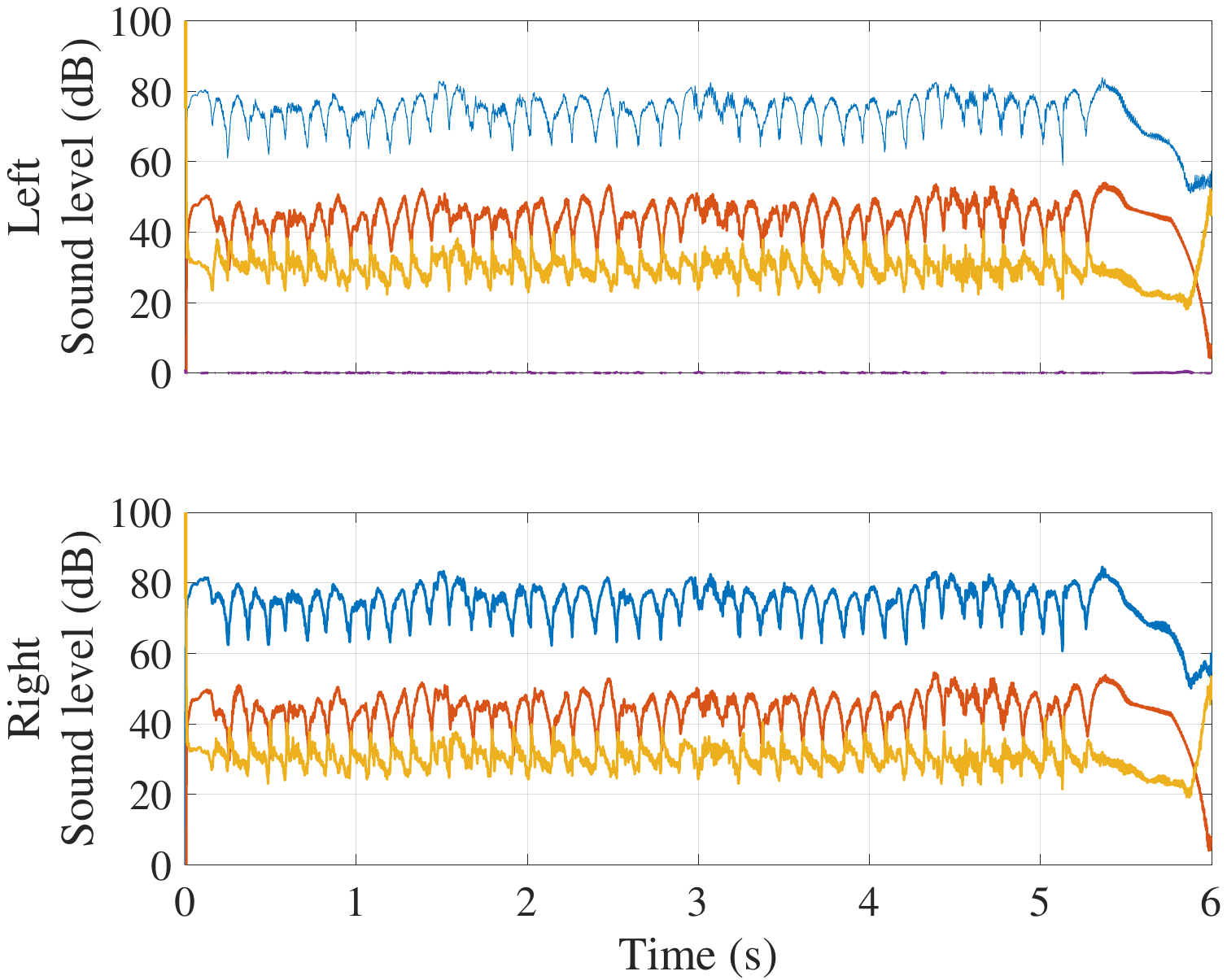} &
\includegraphics[width=0.45\textwidth]{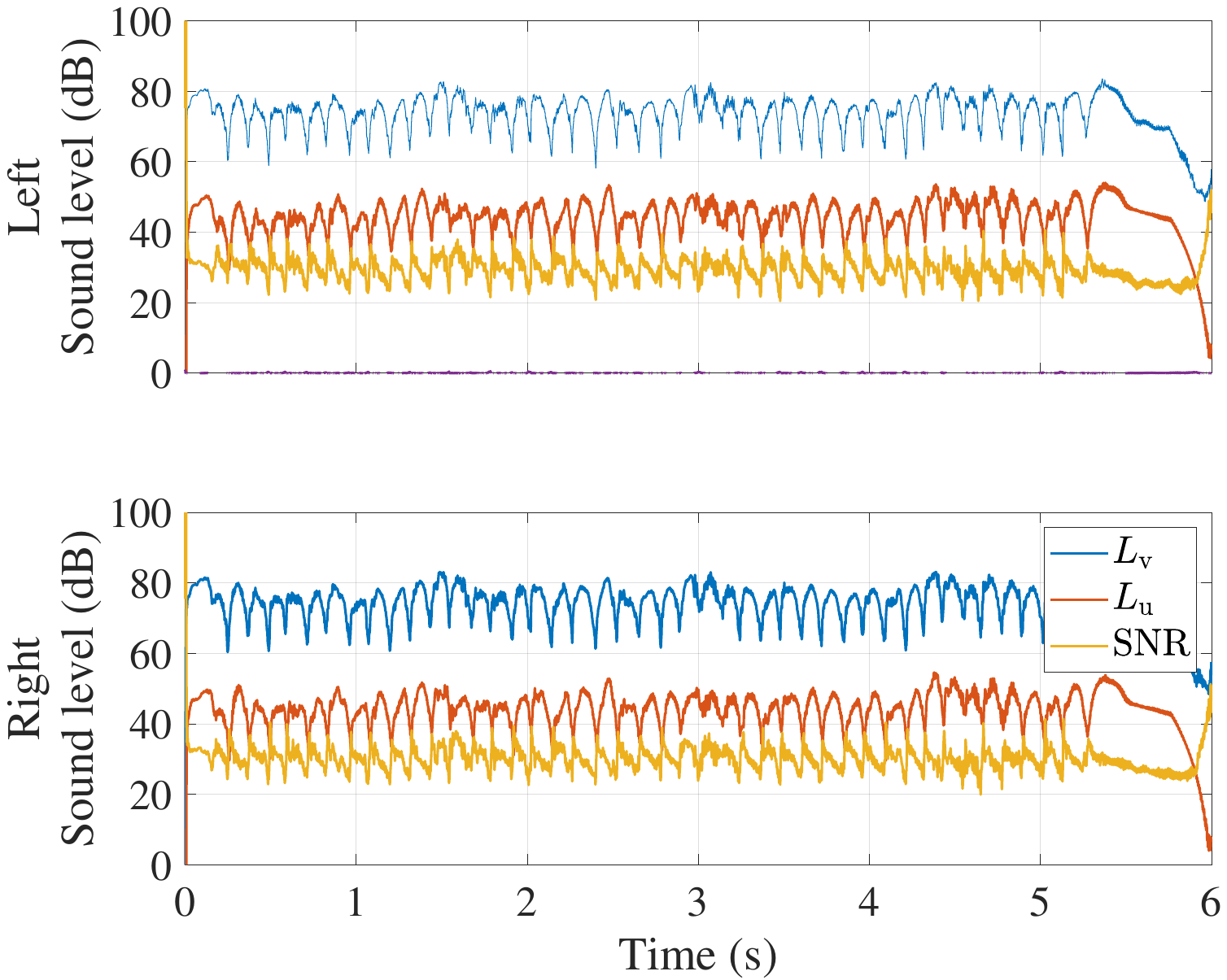} \\
\\
c) Small Virtual Hall -- Hearing booth 1 &
d) Large Virtual Hall -- Hearing booth 1 \\
\includegraphics[width=0.45\textwidth]{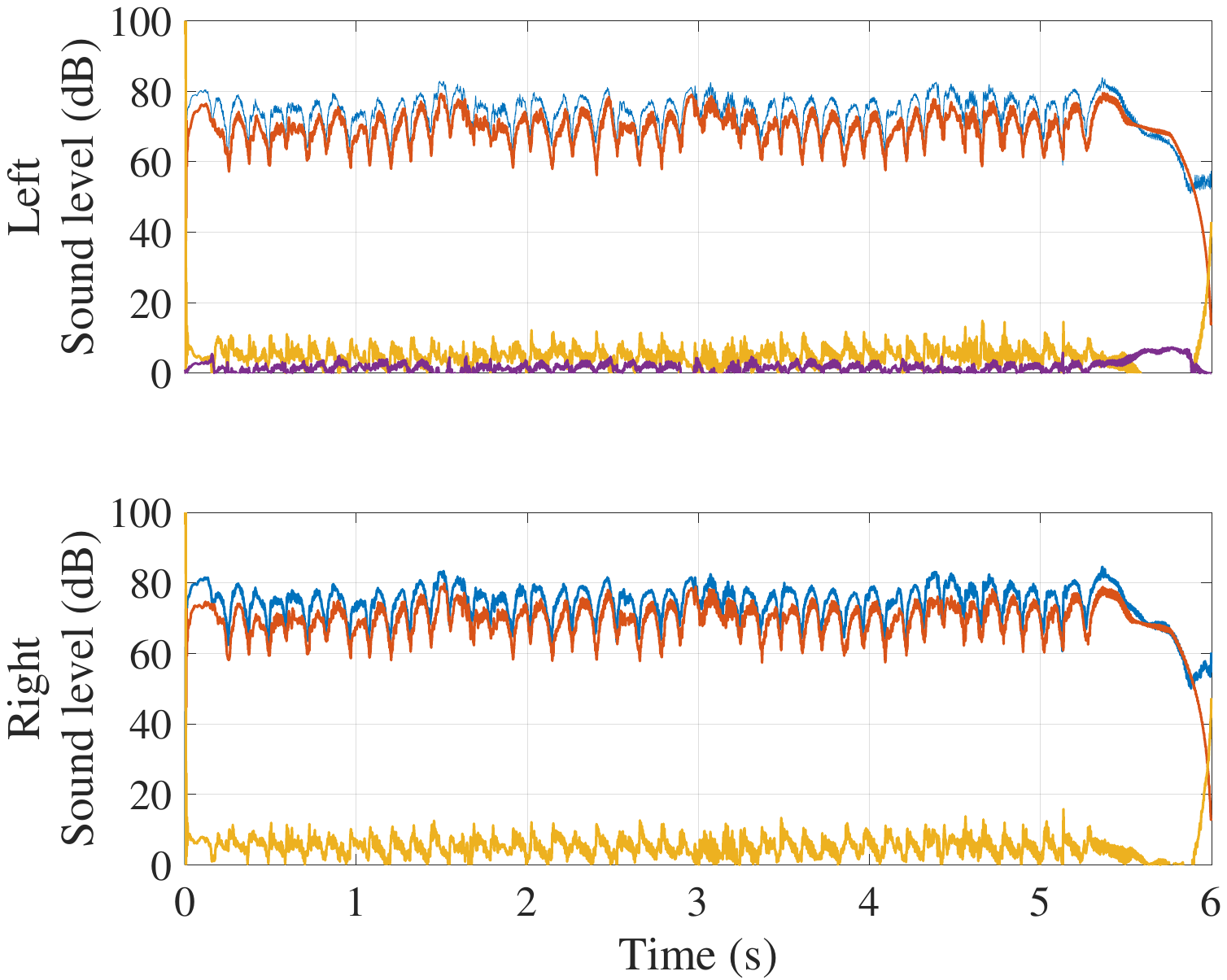} &
\includegraphics[width=0.45\textwidth]{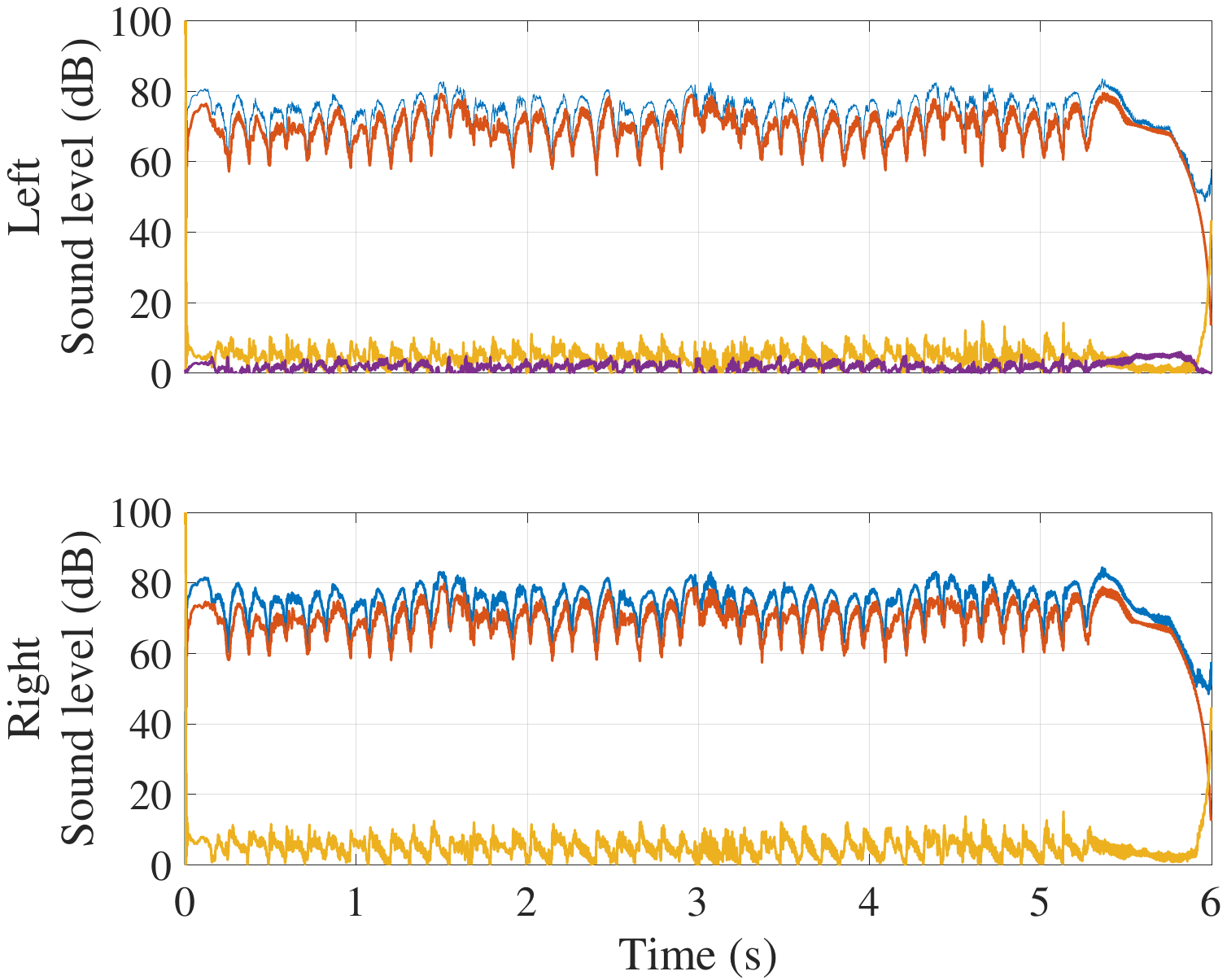} \\
\\
e) Small Virtual Hall -- Hearing booth 2 &
f) Large Virtual Hall -- Hearing booth 2 \\
\includegraphics[width=0.45\textwidth]{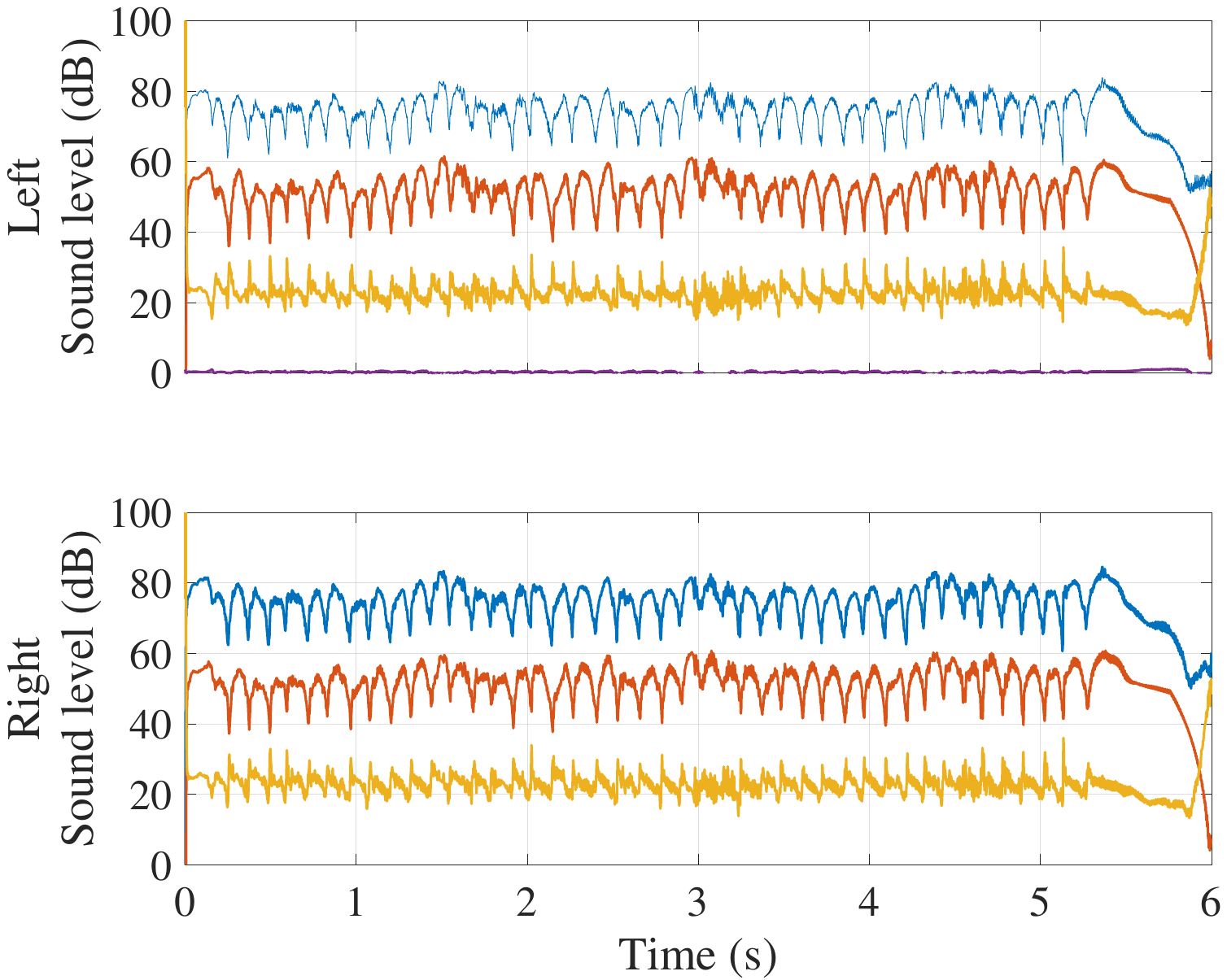} &
\includegraphics[width=0.45\textwidth]{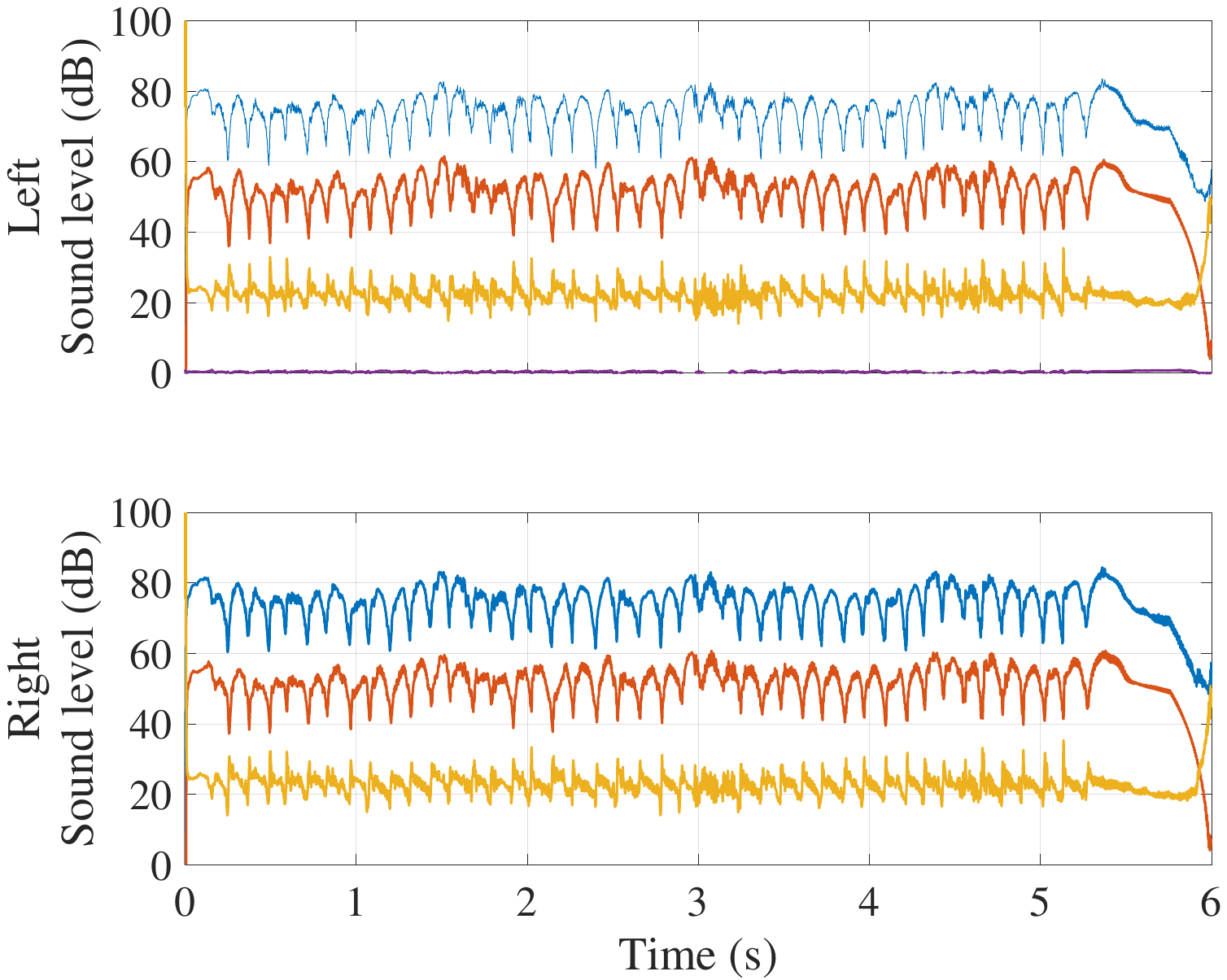} \\
\end{tabular}

\caption{Sound pressure levels comparison for the three laboratory rooms.
The left column (a, c, e) shows the small virtual stage and the right column (b, d, f) shows the large virtual stage.
The top row (a, b) shows the anechoic room, the middle row (c, d) shows the hearing booth 1, and the bottom row (e, f) shows the hearing booth 2.
Blue line: $L_\mathrm{v}$, red line: $L_\mathrm{u}$, green line: $\mathrm{SNR}$. 
}
\label{fig:levels}
\end{figure*}

Fig.~\ref{fig:levels} shows $L_\mathrm{v}$, $L_\mathrm{u}$ and $\mathrm{SNR}$ for the three laboratory rooms and the two virtual stages.
Although $L_\mathrm{v}$  in each column of Fig.~\ref{fig:levels}, $L_\mathrm{u}$ and $\mathrm{SNR}$ vary for each row.
Besides, the signals $y_{\mathrm{v}}$, $y_{\mathrm{u}}$, and $y_{\mathrm{t}}$ are available as supplementary material.

\begin{figure}[h]
\centering
\includegraphics[width=0.5\textwidth]{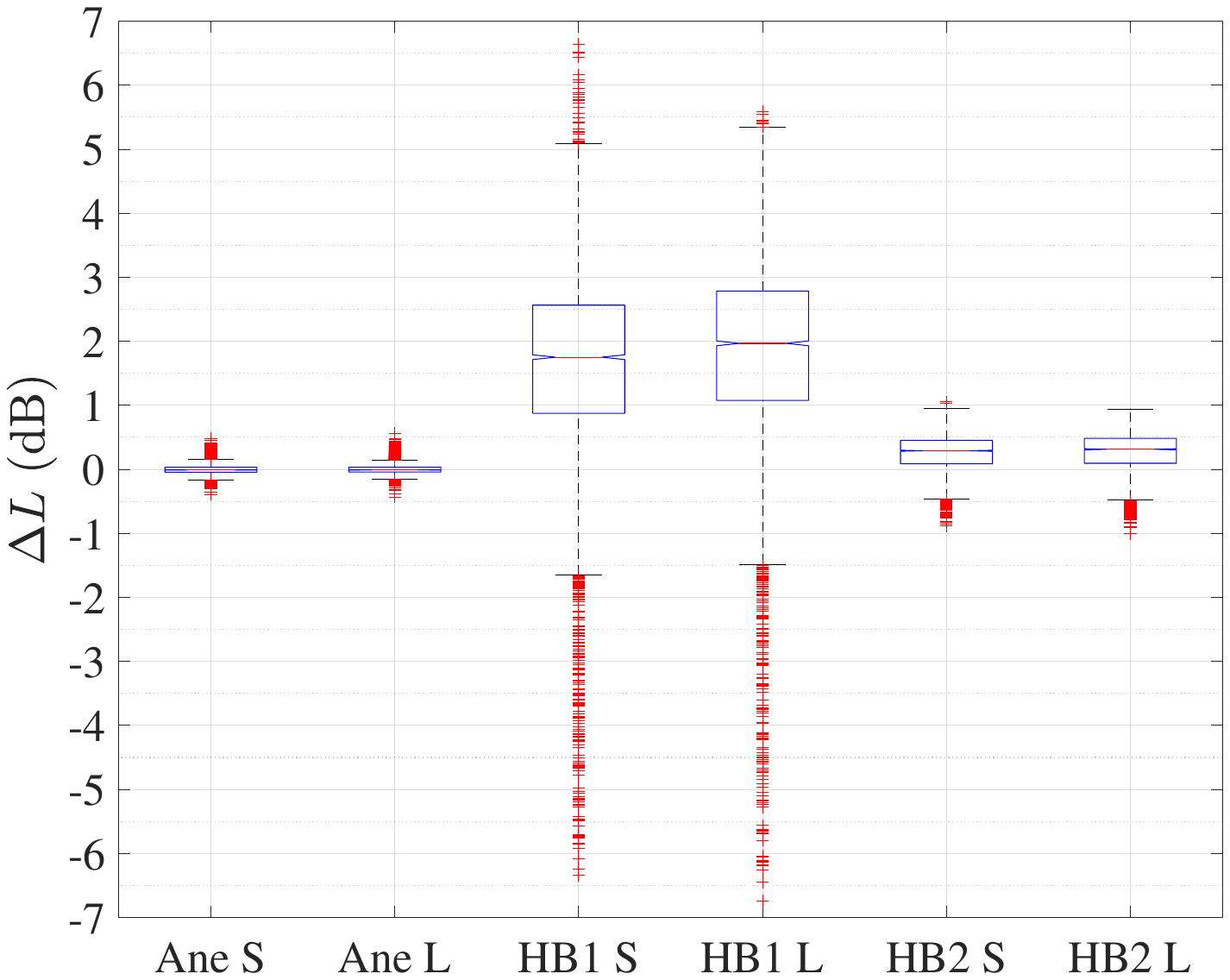}
\caption{Boxplot of $\Delta L$ for the three laboratory rooms and the two virtual stages.}
\label{fig:boxplot}
\end{figure}

Fig.~\ref{fig:boxplot} shows the boxplots of $\Delta L$ for the three laboratory rooms and the two virtual stages.
The box plots the interquartile range (IQR) of data, with the central line representing the median, the box edges indicating the first and third quartiles, and the whiskers extending to 1.5 times the IQR from the quartiles.

\section{Discussion}
\label{sec:discussion}

The residual sound may be assumed as background noise; hence the difference between the virtual sound and the residual sound is a measure of the signal-to-noise ratio of the virtual sound.
The SNR is higher for the anechoic room than for the hearing booths.
In turn, the SNR is higher for the hearing booth 2 than for the hearing booth 1. 

Since the residual sound is actually a copy of the direct sound, low SNRs may cause virtual sounds being listened louder than they should.
This effect may be perceived when $L_\mathrm{t}$ is at least a just-noticeable difference (JND) louder than $L_\mathrm{v}$.     

A JND around 1 dB may be assumed for $\Delta L$ based on the value  stated for sound strength $G$ in ISO 3382-1 \cite{iso3382_1}. 
This parameter $G$ measures the effect of the room perceived at the audience of a concert hall.
Experimental congruent evidence states that the JND for a piano tone is 0.68--1.22 dBC \cite{slade2023}. 

The boxplot in Fig.~\ref{fig:boxplot} shows that the anechoic room has a median $\Delta L$ of 0.01 dB for both the small and large virtual stages, which is below the JND. 
The hearing booth 1 has a median $\Delta L$ of 1.8 dB for the small virtual stage and 2.0 dB for the large virtual stage, which is above the JND.
The hearing booth 2 has a median $\Delta L$ of 0.3 dB for the small and the large virtual stages, which is below the JND. Besides, the 75\% percentile of the hearing booth 2 is 0.45 dB for the small virtual stage and 0.48 dB for the large virtual stage, which is also below the JND.
Considering points outside the 1.5 IQR,  the anechoic room has outliers with absolute value smaller than 0,56 dB, the hearing booth 2 has outliers with absolute value smaller than 1.06 dB, and the hearing booth 1 has outliers with absolute value smaller than 8.75 dB. 

The results show that the anechoic room and the hearing booth 2 have a $\Delta L$ below the JND, which means that the virtual sound is not perceived louder than it should.
In contrast, the hearing booth 1 has a $\Delta L$ above the JND, which means that the virtual sound is perceived louder than it should.
The hearing booth 2 is a compromise scenario, which means that the virtual sound is not perceived louder than it should, but the SNR is lower than for the anechoic room.

Beyond changes in perceived loudness, the presence of residual sound can potentially influence other critical perceptual attributes. However, any

\section{Conclusion}
\label{sec:conclusion}

This preliminary study assessed the impact of the acoustical conditions of three different rooms on the perception of virtual stages for music, providing foundational insights for the design and validation of augmented acoustics laboratories.
The results show that the anechoic room and the hearing booth 2 have a difference between the total sound and the virtual sound below the just-noticeable difference, which means that the virtual sound is not perceived louder than it should.
In contrast, the hearing booth 1 has a difference above the just-noticeable difference, which means that the virtual sound is perceived louder than it should.
The hearing booth 2 is a compromise scenario, which means that the virtual sound is not perceived louder than it should, but the signal-to-noise ratio is lower than for the anechoic room.
This study provides a preliminary validation of the proposed methodology for assessing the acoustical conditions of custom-made hearing booths in stage acoustics experiments, underscoring the importance of rigorous laboratory validation to ensure the perceptual accuracy and reliability of virtual acoustic environments. These findings form a critical basis for future, more comprehensive perceptual investigations and the development of empirically-driven design guidelines for augmented acoustics research facilities.
Future work will include a more comprehensive analysis of the results, including the effect of different musical stimuli, perceptual validation, and the development of robust guidelines for laboratory design in virtual acoustics research.

\bibliographystyle{apalike}
\bibliography{Lab_Stage_biblio}

\begin{thebibliography}{}

\bibitem[Accolti \textit{et al.}, 2023]{accolti2023}
Accolti, E., Aspöck, L., y Vorländer, M. (2023).
\newblock An approach towards virtual stage experiments.
\newblock In {\em Forum Acusticum 2023}, Torino, It.

\bibitem[Ackermann y Brinkmann, 2024]{ackermann2024}
Ackermann, D. y Brinkmann, F. (2024).
\newblock A database with directivities of musical instruments.
\newblock {\em J. Audio Eng. Soc.}, 72(3).

\bibitem[Braak \textit{et al.}, 2019]{braak2019}
Braak, E. V.~D., Marshall, H., Meyer, J., Halstead, M., y Protheroe, D. (2019).
\newblock Further investigation of ensemble singers preferred sound fields.
\newblock In {\em International Symposium on Room Acoustics (ISRA 2019)}.

\bibitem[Brinkmann \textit{et al.}, 2017]{brinkmann2017}
Brinkmann, F., Lindau, A., Weinzerl, S., Van De~Par, S., Müller-Trapet, M.,
  Opdam, R., y Vorländer, M. (2017).
\newblock A high resolution and full-spherical head-related transfer function
  database for different head-above-torso orientations.
\newblock {\em Journal of the Audio Engineering Society}, 65(10):841--848.

\bibitem[Chalupper y Fastl, 2002]{chalupper2002}
Chalupper, J. y Fastl, H. (2002).
\newblock Dynamic loudness model (dlm) for normal and hearing-impaired
  listeners.
\newblock {\em {Acta Acustica united with Acustica}}, 88(3):378–386.

\bibitem[Gade, 1989]{gade1989}
Gade, A. (1989).
\newblock Investigations of musicians’ room acoustic conditions in concert
  halls. part i. methods and laboratory experiments.
\newblock {\em Acustica}, 69(5):193–203.

\bibitem[{ISO 3382-1}, 2009]{iso3382_1}
{ISO 3382-1} (2009).
\newblock Acoustics - {Measurement} of room acoustic parameters - {Part} 1:
  {Performance} spaces.

\bibitem[Moore, 2014]{moore2014}
Moore, B. C.~J. (2014).
\newblock { Development and Current Status of the “Cambridge” Loudness
  Models}.
\newblock {\em Trends in Hearing}, 18.

\bibitem[Pausch, 2022]{pausch2022}
Pausch, F. (2022).
\newblock Documentation of the experimental environments and hardware used in
  the dissertation ``{Spatial} audio reproduction for hearing aid research:
  System design, evaluation and application''.
\newblock Technical report, RWTH Aachen University.

\bibitem[Schröder y Vorländer, 2011]{raven}
Schröder, D. y Vorländer, M. (2011).
\newblock Raven: A real-time framework for the auralization of interactive
  virtual environments.
\newblock In {\em Forum Acusticum}, pp. 1541--1546.

\bibitem[Slade \textit{et al.}, 2023]{slade2023}
Slade, T., Gascon, A., Comeau, G., y Russell, D. (2023).
\newblock Just noticeable differences in sound intensity of piano tones in
  non-musicians and experienced pianists.
\newblock {\em Psychology of Music}, 51(3):924–937.

\bibitem[Thery y Katz, 2019]{thery2019}
Thery, D. y Katz, B. F.~G. (2019).
\newblock Anechoic audio and 3d-video content database of small ensemble
  performances for virtual concerts.
\newblock In {\em International Congress on Acoustics (ICA~2019)}.

\end{thebibliography}

\end{document}